# Using Data Science to monitor the pandemic with a single number: the Synthetic COVID Index

Raffaele Zenti*

*Virtual B SpA, Milan, Italy*
raffaele.zenti@virtualb.it

*Abstract*— Rapid and affordable methods of summarizing the multitude of data relating to the pandemic can be useful to health authorities and policy makers who are dealing with the COVID-19 pandemic at various levels in the territories affected by SARS-CoV-2. This is the goal of the Synthetic COVID Index, an index based on an ensemble of Unsupervised Machine Learning techniques which focuses on the identification of a latent variable present in data that contains measurement errors. This estimated latent variable can be interpreted as "the strength of the pandemic". An application to the Italian case shows how the index is able to provide a concise representation of the situation.

I. INTRODUCTION

In each country, numerous data are collected related to the expansion of the SARS-CoV-2 epidemic, e.g., reported cases, notification rates, deaths, hospitalisation and ICU, testing capacity, and so on.

All these numbers - highlighted daily by the media in a frenzied flurry of news - reflect the strength of the contagion, but no one can fully represent the situation on its own. Furthermore, the various metrics related to COVID-19 are often plagued by errors of various kinds, such as counting errors, software and human errors, including data entry problems, delays in the calculations of cases: that is, a variety of situations caused by a set of factors related to the effectiveness of health surveillance systems. It is a widespread belief that, if data are flawed, sound public policy decisions cannot be made; therefore, it is important to have indicators designed to explicitly take into account the fact that data contain measurement errors. And this is what the method proposed in this work does.

Furthermore, when having to establish suitable policies to deal with COVID-19, on one hand it is paramount to have an analytical and detailed framework, and on the other it is useful to have summary indicators, which can provide a compact representation of the situation. An indicator of this type was built by [3] using the Balanced Worth methodology proposed by [6], a completely different approach from that proposed in this study, which instead has its foundations in Data Science.

The objective of this paper is to present a methodology based on an ensemble of Unsupervised Machine Learning techniques, focusing on the identification of latent variables present in the data associated with various phenomena. These methods, widely used in the field of Data Science are able to provide a data-driven, concise overview - literally a single number per day, the Synthetic COVID Index - of the epidemiological situation of the COVID-19 pandemic in a given region, country or continent, keeping into account and reducing the effect of errors in the data. Such an index can be a useful complement to more detailed indicators to support health authorities and policy makers dealing with the pandemic at various levels in the territories affected by COVID-19.

The structure of this paper is as follows: in section II the methodology is presented, in section III a simple example will be shown - an application to the Italian situation. Finally, in section IV some concluding remarks will be made.

## II. KEY IDEAS AND METHODS

From a methodological point of view, the two main ideas behind the Synthetic COVID Index are the use of a particular class of Unsupervised Machine Learning models, that is, latent variables models, and the application of the model averaging principle, which leads to the use of an ensemble model.

### A. Latent variable modelling

In many fields, a frequently encountered issue is the analysis of data to extract meaningful latent, not observable factors from it.

The idea of latent variables arises from the observation of the following phenomenon: in datasets with many variables, groups of them often move together because they might be reflecting the same driving principle (or principles) governing the behaviour of the system. In many systems there are only a few such latent variables, but there is a certain abundance of observable variables (i.e., manifest variables). When this happens, it is possible to take statistical advantage of this redundancy of information, as one can try to infer the latent variables using a latent variable model.

A latent variable model mathematically relates a set of observable variables to a set of latent variables. By using the model, the values of the latent variables can be inferred from measurements of the observable variables. Thus, latent variable models are aimed to reveal "the underlying truth" that hides behind an observed and measured phenomenon, in the presence of measurement errors.

In many practical situations it is interesting to simply grasp the main effect, which often explains a large part of the phenomenon, albeit with approximation. So, the focus is on one, or two latent variables clearly present in the dataset. In this specific application, for example, the focus is on one single latent variable, i.e., the main driver of the pandemic. In this way the problem can be simplified by replacing a group of variables with a single new variable (and in fact latent variable models are widely used for the purpose of dimensionality reduction in Machine Learning and Artificial Intelligence applications).

To grasp the idea of a latent variable model, a very instructive example is provided by State-Space models (see [4] and [19]).

A State-Space model is a linear, discrete-time, stochastic model that describes a multivariate system:

$$x(t) = A(t) \cdot x(t-1) + B(t) \cdot \xi(t) \quad (1.1)$$
$$y(t) = C(t) \cdot x(t) + D(t) \cdot \varepsilon(t) \quad (1.2)$$
$$t = 1,2,3,\ldots,T$$

where:

$x(t)$ is a $m$-dimensional "state vector" describing the state of some unobservable phenomenon at time $t$, i.e., the latent variables;

$y(t)$ is a $n$-dimensional "measurement vector" with the values of some observable metrics at time $t$;

$A(t)$ is a $m \times m$ state-transition matrix, that describes the deterministic component of the evolution of $x$ between $t-1$ and $t$;

$B(t)$ is a $m \times k$ state-disturbance-loading matrix, that take into account random disturbances, i.e., $\xi$;

$\xi(t)$ is a $k$-dimensional vector of random state-disturbances;

$C(t)$ is a $n \times m$ measurement-sensitivity matrix, that describes how the observations at time $t$ relate to the state vector at time $t$;

$D(t)$ is a $n \times h$ measurement-disturbance-loading matrix, that considers random observation errors, i.e., $\varepsilon$;

$\varepsilon(t)$ is a $h$-dimensional vector of random measurement errors; $\varepsilon(t)$ and $\xi(t)$ are orthogonal, i.e., uncorrelated, and can follow any probability distribution, even if for computational reasons they are usually assumed to be Gaussian, white-noise, unit-variance vectors.

Thus, a State-Space model contains two sets of equations: the "state equations" group (1.1) describes how a latent process transitions in time, while the "observation equations" group (1.2) describes the dynamics of the measured variables. Both sets of equations contain random disturbances, and therefore reflect random elements in the process and measurement errors. This is exactly the point of interest when trying to cope with noisy, messy COVID data. In addition, the coefficient matrices $A$, $B$, $C$, $D$ might change from period to period ("Time-

Varying State-Space model"), or they can be time-invariant ("Time-Invariant State-Space models") reflecting structural changes in the system. The information above illuminates how such a framework is suitable for capturing phenomena:
- that vary over time;
- disturbed by errors and stochastic factors;
- whose evolutionary structural laws (i.e., the coefficient matrices) may vary over time.

The brief overview on the State-Space models was due to their flexible structure, and their adaptability to many situations. These models lend themselves well to presenting the logic of latent variable modelling in general, and they facilitate the understanding of why latent variable models are suitable for measuring the evolution of the pandemic in a synthetic way.

There are many different types of latent variable models. At a first level of analysis, they all have the same objective - that is, to deal with latent variables - but they can present significant differences as soon as one goes down to a level of greater detail. They differ, for example, according to whether the observable and latent variables are continuous or categorical, if parameters can vary or not, if the underlying data generation process is linear or not, if it is Gaussian or not. The details and an in-depth discussion of these models are beyond the scope of this paper (for a rather comprehensive overview of statistical latent variable models, see [1]), which merely and briefly present the most popular ones, used in the example of the next section.

COVID data are typically noisy time series of variables that can be safely considered continuous, e.g., daily cases, deaths, number of tests performed. Table 1 lists some rather popular latent variable models that might be used for modelling this kind of data.

TABLE I
POPULAR LATENT VARIABLE MODELS

| Model | Short description |
|---|---|
| Principal Component Analysis (PCA) | Each PC is a linear combination of the original variables. There is no redundant information, as PCs are uncorrelated to each other (they form an orthogonal basis for the space of the data). PCA uses up to the second order moment of the data to produce uncorrelated components, so it is well suited when data follow a Gaussian distribution. |
| Independent Component Analysis (ICA) | ICA is a generative model that estimates components as independent as possible, minimizing higher order dependencies, i.e., relaxes the Gaussian hypothesis underlying PCA. From other practical points of view, ICA is quite similar to PCA |
| Common Factor Analysis (CFA) | In the CFA model there are some measured variables that depend on a smaller number of unobserved latent factors. Each observed variable is assumed to depend on a linear combination of the common factors, and is impacted by some independent random variability, specific to each single measured variable. CFA seeks the least number of factors which can account for the common variance (correlation) of a set of observable variables. |
| Canonical Factor Analysis (CanFA) | Very similar to CFA, CanFA seeks latent factors which have the highest canonical correlation with the observed variables. |
| State-Space Models (SSM) | A SSM is a system of first-order difference equations characterized by state and observation equations, usually estimated by the Kalman Filter, a Bayesian recursive filter for multivariate normal distributions that estimates the values of the latent stochastic variables, based on possibly mismeasured observations. |
| Sparse Filtering (SF) | Instead of explicitly attempting to construct a model of the data distribution, SF optimizes a simple cost function - normalized sparsity penalty, i.e., the sparsity of L2-normalized features. It is essentially a hyperparameter-free model. |
| Cointegration Models (CM) | Cointegration deals with the presence of "common trends" (i.e., latent variables) that determine common behaviour of multivariate time series, "the cointegrating systems". A CM is a representation of a cointegrating system - there are several different approaches. |

For an overview of Data Science methods see [2] and [5]. On PCA, CFA and other methods based on matrix factorization see [8] and [9]; specifically, on ICA, see [15]. See [10] on SF, and [12] on CM.

*B. Ensemble modelling*

The previous paragraphs illuminated how models for latent variables can explicitly account for measurement errors - see for example the State-Space model represented by equation (1.1) and (1.2.). Unfortunately, there is also the specification error: the model can be structurally wrong. As one does not know what the truth is, there is structural model uncertainty.

On a conceptual level, if reality is represented by $M$, the "true model", it is possible to use the model $\widehat{M}$ instead. This model exhibits a specification error $\varsigma \geq 0$ of unknown size - which has the nature of a non-negative random variable (i.e., in the best case the error is zero, which means that one knows the truth), that is:

$$\widehat{M} = M + \varsigma \qquad (2)$$

It is hard to overstate the importance of model uncertainty in quantitative modelling, especially when using Unsupervised Machine Learning models, that work without target variables: almost invariably empirical work in this area is subject to a large amount of uncertainty about model specifications. Taking a 'leap of faith' and narrowly focusing on a limited set of possible models or, worse, on a single model, implies that one will likely fail to capture important aspects of reality.

In particular, if the goal is summarizing the strength of the pandemic, a quintessentially unobservable variable, it is quite unlikely that reality can be adequately captured by any single model, and it would be risky to rely on a single selected model for inference and policy decisions of vital interest. It should always be kept in mind that modelling assumptions may be wrong, and wrong modelling assumptions may determine wrong conclusions: this can lead to dramatically incorrect policy choices, especially when talking about a pandemic. Furthermore, one cannot forget that the specification error is added to the measurement error, so the number of estimation problems that can be encountered in this situation is high.

Here are some concrete examples of model uncertainty. First, one can be uncertain about the nature of latent variables, their number and (maybe) even their existence. There is the problem of weak identifiability - a model is non-identifiable if distinct parameterizations lead to identical data distributions (i.e., similar results) - and this is often the case with Unsupervised Machine Learning models. Uncertainty concerns both the type of model (for example, the researcher is unsure about the functional form of the models, State-Space, or CFA, or ICA) and its hyperparameters (for example, the researcher is unsure about covariate inclusion, or a data window for a Rolling ICA, or about the number of features to extract from the sample data using Sparse Filtering).

Since "seeing is believing", the situation is aggravated by the fact that when modelling unobservable latent variables there are no response variables, so the check is difficult, if not impossible. And interpretation may be fairly ambiguous. In most cases, only use and experience help evaluating Unsupervised Machine Learning models of this kind.

Model averaging is a (partial) solution to the problem of structural model uncertainty, i.e., specification error, and it can be justified in various ways. For example, on a completely conceptual level, if the specification error $\varsigma$ in the model space has just finite mean $\mu$ and standard deviation $\sigma$ (without any additional hypothesis about the shape of its probability distribution) using the Cantelli inequality one can say that (see [13]):

$$Prob(\varsigma - \mu \leq -\alpha \cdot \sigma) \leq \frac{1}{1+\alpha^2}, \quad \alpha > 0 \qquad (3)$$

This inequality helps to understand how much a single realization of the specification error - i.e., a specific model in which "one believes" - can be separated from its mean (i.e., the result of model averaging). According to this probability inequality, the probability that a single model has a specification error of, say, 2 standard deviations lower than the model mean is 20%, of 3 standard deviations is 10%. That is, it is not very likely that a single model drawn from the universe of possible models is significantly better than an average of models.

Similarly, with the additional hypothesis that the probability distribution of the error is symmetric, from the Bienayme-Tchebycheff inequality (see [13]) it follows that:

$$Prob(\varsigma - \mu \leq \lambda \cdot \sigma) \leq \frac{1}{2 \cdot \lambda^2}, \quad \lambda > 0 \quad (4)$$

The slightly stricter requirement on the probability distribution of the specification creates sharper conditions, and, for example, the possibility that a single model has a specification error of 2 standard deviations lower than the model mean is 12.5%, of 3 standard deviations is about 5.5%.

Now, these are approximate arguments: nobody knows the probability distribution of the specification error, no one is sure that the empirical average of a finite number of models is a good estimator of $\mu$ (even if one can rely on the Weak Law of Large Numbers to hope that the sample average of their models converges in probability towards $\mu$), nor it is possible to know the standard deviation $\sigma$ (but if the model space has been conscientiously chosen, it is likely to be rather low). Therefore, summing up, under broad conditions, the probability that model averaging leads to better results than a single specific model is rather sensible.

Furthermore, model averaging has its roots solidly in Bayesian Decision Theory. Indeed, in line with probability theory, the Bayesian response to dealing with uncertainty, in general, is to take a (smart) average. When dealing with parameter uncertainty, in order to get the predictive distribution, this implies averaging over parameter values using the posterior distribution of that parameter. In a completely analogous way, model uncertainty is addressed by the Bayesian statistics through averaging, but this time averaging over models, with the discrete posterior model distribution.

Based on the information reviewed in the previous paragraphs, it is possible to say that averaging models theoretically makes sense, supporting the direct intuition that different models, based on different concepts, allow for looking at the "physical truth" from different angles, or to use contrasting approaches to prediction. Lastly, anecdotal evidence shows that in practice an average of models frequently performs better than any individual model, because the various errors of the different models "average out".

The broad idea of combining models can be applied in different ways: it is possible both to combine the models themselves, and to combine the forecasts (i.e., the outcomes of the models). Common types of ensembles in Machine Learning are Bootstrap aggregating, Bayes optimal classifier, Boosting, Bayesian model averaging, Consensus Clustering, just to name a few. However, the principle of model averaging is very general: a large range of different model averaging methods exists, but a detailed discussion on this complex and broad topic goes beyond the scope of this article - see [11] for an overview.

The typical result of the model averaging is a generic model ensemble, i.e., alternatively:
- multiple models combined to produce a desired output, that is a (hopefully better) final prediction;
- multiple outputs combined to hopefully get a better result than the output of a single model.

Therefore, models or predictions are averaged over all the considered models, using weights that are either derived from Bayes' theorem, or other frequentist methods.

*C. Meta-algorithm*

Summing up, the algorithm used to create the Synthetic COVID Index combines the outputs of different latent variables models.

In practical applications, a key aspect of dealing with model uncertainty is the definition of the space of all models that are being considered.

In principle, the idea of model averaging naturally assumes a well-defined space of possible models, over which the averaging takes place. In the real world, things are often not that clear, especially with regards to Unsupervised Machine Learning methods and latent variable models, thus modelling experience and "statistical common sense" should be used in choosing the model space.

To give an example, a Time-Varying State-Space model in principle gives a better description of a dynamically evolving phenomenon, however it has a large number of parameters, the estimate of which absorbs a lot of data. So, if the sample is not very large, the variance of the estimates can be very high. Conversely, a Time-Invariant State-Space model may offer a worse description of the phenomenon, but it leads to more robust estimates. Therefore, one model is dominated by measurement error, the other by specification error; it is very likely that by including both types of models, the model risk diversifies, approaching an optimal trade-off. To give another example, the PCA applied over the entire period offers a static and crystallized vision of the phenomenon (which is essentially summarized in the eigenvector corresponding to the first component). Nonetheless, the estimates are likely to contain less measurement error than a Rolling PCA (i.e., using a shorter, forward moving window) which enables one to do a kind of time-dependent calculation. Yet, using less sample data might contain more measurement error.

Summing up, the model space should be well-matched both in terms of different types of models and their hyperparameters, trying to cope with model risk by balancing specification risk and measurement risk. Predictions of different models are averaged over all the considered models, using an equal weighting scheme that corresponds to the Bayesian optimal solution for combining the different outputs in the absence of specific information on the quality of the fit and the predictive goodness of the models - which is logical, since there are no target variables.

Needless to say, before the estimation the raw data are treated with standard techniques to fill in any missing values, to reduce the noise and to standardize (computing z-scores). The result is rebased (i.e., it starts at 100).

The meta-algorithm for the Synthetic COVID Index is as follows:

## META-ALGORITHM

**Input**: Data = matrix $n_{observations} \times n_{input\ variables}$

**Output**: Synthetic COVID Index = vector $n_{observations} \times 1$

- fill empty values (e.g., with the EM algorithm), standardize, and reduce daily noise in Data (e.g., using any low-pass filter such as exponential smoothing, or moving average, or ARMA, etc);

- define the $K$ types of models (macro-models) $Model_i, i = 1, \ldots, K$, to include in the model space (e.g., PCA, SSM, ICA, etc.);

- for each type of model $Model_i$ define all the $H_i$ combinations of suitable hyperparameters (micro-models), so the model space has size $S = \sum_i H_i$, which means that one has $S$ micro-models;

- estimate each of the $S$ micro-models as a function of Data (transforming data as needed, e.g., taking first differences, logs, using rolling windows, etc.);

- use each of the $S$ micro-models to obtain an estimate of the main latent variable $X(t)$ (e.g., first component of a PCA, first factor of a CFA, hidden "filtered" factor of a SSM, etc), thus obtaining $S$ estimates of the latent variable, $X(t)_j, j = 1, \ldots, S$;

- compute the Synthetic COVID Index = $\overline{X(t)} = \frac{\sum_j^S X(t)_j}{S}$ (and properly rebase it, e.g., make it start at 100).

It must also be said that the amount of input variables might change over time: the algorithm can accept a different data matrix in input.

### III. AN EXAMPLE: THE ITALIAN CASE

As an application example, the Synthetic COVID Index for Italy, at the national level, was estimated. The data used are the official ones of the Italian Government / Protezione Civile and are public (see [7]), available from 2020-02-24 on a daily basis.

The raw variables taken into consideration on a daily basis are the following:
- Total amount of positive cases;
- Hospitalised patients with symptoms;
- Patients in intensive care;

- People in home confinement;
- Deceased people;
- Tests performed.

In the context of the various hyper parameterizations of latent variable models, both the data in original form and various transformations of such data are considered (e.g., absolute variation rates, percentages, logarithmic first differences, etc.). Obviously, if other data were available, in addition to, or in place of these, it would not be an issue: the strength of the latent variable models is precisely that of 'digesting' multitudes of different observable measures and extracting information on the underlying phenomenon (i.e., the latent factor).

Figure 1 shows some of the series used as inputs: the effect of (at least) a common latent variable, linked to the spread of the infection, is intuitively rather evident.

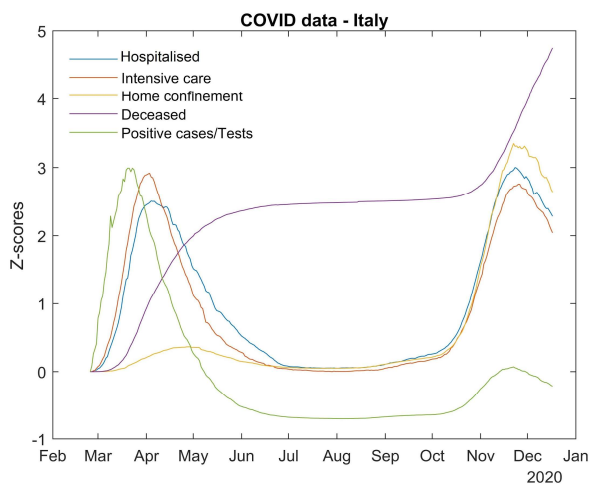

Fig. 1 - A plot of the time series of the z-scores of the input data: hospitalised patients with symptoms, patients in intensive care, people in home confinement, deceased people, total amount of positive cases/tests performed (period 02/2020 – 12/2020). . Data is denoised using a low-pass filter.

Figure 2 shows the Synthetic COVID Index for Italy from the beginning of the pandemic (February 2020).

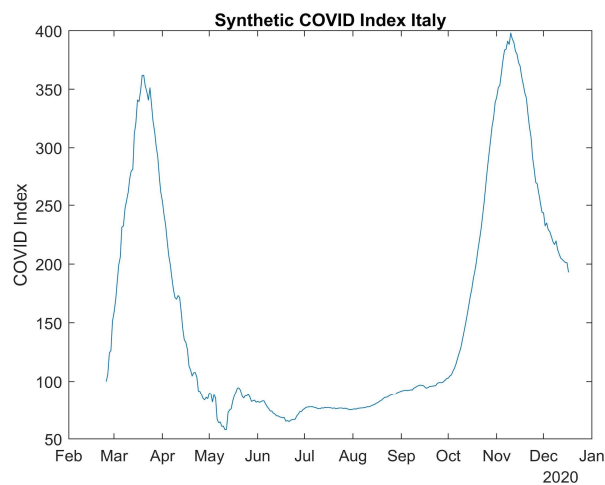

Fig. 2 - The Italian Synthetic COVID Index, for the period 02/2020-12/2020.

The index shows two clear peaks in the spring, and in the fall. The impacts are visible: just observe, for example, the co-movements of the Synthetic COVID Index with the main Italian Stock Exchange index (FTSE MIB) - see Figure 3.

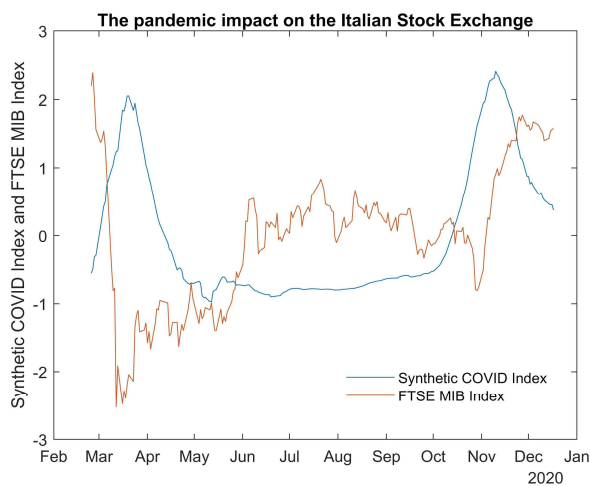

Fig. 3 - The Italian Synthetic COVID Index, and the most popular Italian stock market index, the FTSE MIB, for the period 02/2020-12/2020. Both indices are scaled using their z-scores.

The presence of index co-movements with economic and financial indicators such as the FTSE MIB index indicates its possible usefulness as a general 'thermometer of the situation' also to those involved in financial investments.

When analysing the dynamics of the Synthetic COVID Index together with an index of spread of the infection among the 21 regions included in the

dataset, the COVID Spread Index (see Figure 4), defined as follows:

$$CSI_t = \frac{100 \sum_u f_{u,t} \cdot \log_2(f_{u,t})}{\log_2(21)} \quad (5)$$

where $CSI_t$ is the Covid Spread Index at time $t$, $f_{u,t}$ is the relative frequency of new positive cases at time $t$ of the region $u, u = 1,\ldots 21$ compared to the national total at time $t$. Basically, $CSI_t$ is the classic Shannon entropy (see [14], in units of bits, normalized in the range [0,100].

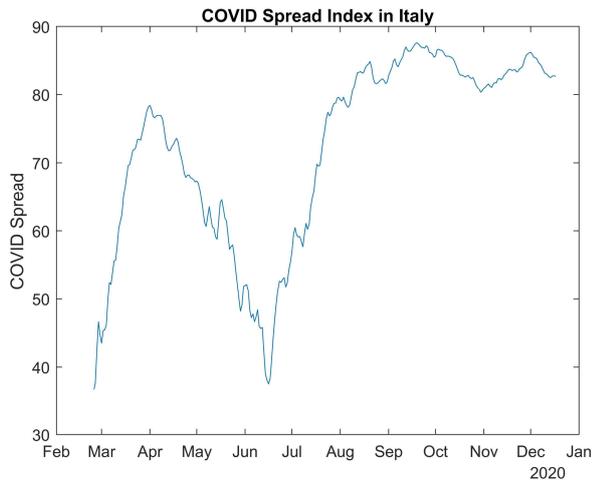

Fig. 4 - The Italian COVID Spread Index, for the period 02/2020-12/2020 (the minimum value is 0, the maximum value is 100).

A scatter plot of the Synthetic COVID Index and the COVID Spread Index - see Figure 5 - shows the relationships between severity and spread of the infection. This helps to understand at a glance the seriousness of the situation distinguishing when:
1. The contagion is strong, but geographically concentrated;
2. It is strong, and widespread on the territory;
3. It is weak, but widespread;
4. It is weak, and concentrated.

Public interventions (e.g., health, economic, and social support) are very different in the four situations highlighted by the joint use of the Synthetic COVID Index and the COVID Spread Index.

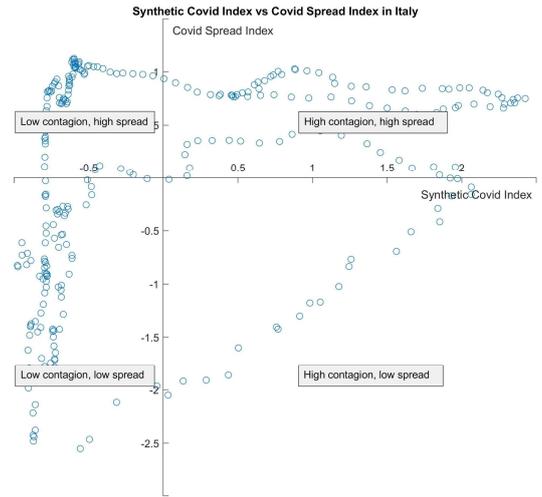

Fig. 5 - A scatter plot of the Synthetic COVID Index vs the COVID Spread Index in Italy, for the period 02/2020-12/2020, shows the relationships between severity and spread of the infection.

Based on the availability of data, once a methodological standard has been defined in its construction the index can be used at the local, regional, national, and continental level, creating a methodologically homogeneous control dashboard (for example, in Italy, the same types of data used here are also available at regional and provincial level).

This common framework could be useful for governments and local administrations to define coordinated interventions.

IV. CONCLUSION AND DISCUSSION

The objective of this work is to present the Synthetic COVID Index, built using a Data Science approach, aimed at effectively synthesizing a multiplicity of data associated with COVID, available at a high frequency (typically daily).

By exploiting the theoretically and practically well-founded idea of model averaging, the index is a structurally robust ensemble of latent variable models. Latent variable modelling encompasses a broad range of statistical and Machine Learning techniques that may be useful for modelling in this context, as they have several desirable properties:
- They acknowledge measurement errors, common in COVID-related data;

- They summarize multiple measures parsimoniously.
- They are well known and widely used models in many fields.

Model averaging is a remedy for the problem of structural model uncertainty. Although specification error can occur with any sort of statistical model, some models and estimation methods are much more affected by it than others. Unfortunately, latent variable models, lacking target variables, are particularly delicate from this point of view. Using weighted averages of several latent variable models, it is possible to reduce specification error, better reflecting model selection uncertainty (as well as reducing prediction error). Model averaging has no super-powers, but it is statistically meaningful and well grounded, and can lead to more robust estimates.

The Synthetic COVID Index can be estimated with a wide variety of data, at various aggregation levels on a geographical basis: local, regional, national, continental. Being a summary metric of the evolution of the pandemic, it can be used, together with other aggregate data (e.g., the COVID Spread Index), as a complement to more analytical data. The Synthetic COVID Index can be a decision support tool for governments and local administrations trying to control and manage the spread of the disease.

Moving forward and thinking about future work, although the Synthetic COVID Index was designed as a "thermometer of the situation", the idea of using it as a short-term predictive tool might be explored.

ACKNOWLEDGMENT

The author wishes to acknowledge Elisabetta Villa, Francesca Cavallerio and Nicola Bedin for the useful insights and constructive discussions on the ideas presented in this article.